\documentclass[a4paper,fleqn]{cas-dc}

\usepackage[numbers]{natbib}
\usepackage{color}

\def\tsc#1{\csdef{#1}{\textsc{\lowercase{#1}}\xspace}}
\tsc{WGM}
\tsc{QE}
\tsc{EP}
\tsc{PMS}
\tsc{BEC}
\tsc{DE}

\begin{document}
\let\WriteBookmarks\relax
\def\floatpagepagefraction{1}
\def\textpagefraction{.001}
\shorttitle{AI/ML methods for coupled folding and binding in disordered protein ensembles}
\shortauthors{A. Ramanathan et~al.}

\title [mode = title]{Artificial intelligence techniques for integrative structural biology of intrinsically disordered proteins}



\author[1,2]{Arvind Ramanathan}
\fnmark[1]
\ead{ramanathana@anl.gov}
\ead[URL]{https://ramanathanlab.org}
\address[1]{Data Science \& Learning Division, Argonne National Laboratory, Lemont, IL, 60439, United States}
\cormark[1]
\address[2]{Consortium for Advanced Science and Engineering (CASE), University of Chicago, Hyde Park, IL, United States}

\author[1]{Heng Ma}


\author[3]{Akash Parvatikar}

\author[3]{S. Chakra Chennubhotla}
\address[3]{Department of Computational and Systems Biology, University of Pittsburgh, Pittsburgh, PA, 15260, United States}

\cortext[cor1]{Corresponding author}

\begin{abstract}
We outline recent developments in artificial intelligence (AI) and machine learning (ML) techniques for integrative structural biology of intrinsically disordered proteins (IDP) ensembles. IDPs challenge the traditional protein structure-function paradigm by adapting their conformations in response to specific binding partners leading them to mediate diverse, and often complex cellular functions such as biological signaling, self organization and compartmentalization. Obtaining mechanistic insights into their function can therefore be challenging for traditional structural determination techniques. Often, scientists have to rely on piecemeal evidence drawn from diverse experimental techniques to characterize their functional mechanisms. Multiscale simulations can help bridge critical knowledge gaps about IDP structure function relationships - however, these techniques also face challenges in resolving emergent phenomena within IDP conformational ensembles. We posit that scalable statistical inference techniques can effectively integrate information gleaned from multiple experimental techniques as well as from simulations, thus providing access to atomistic details of these emergent phenomena.  
\end{abstract}

\begin{graphicalabstract}
\includegraphics{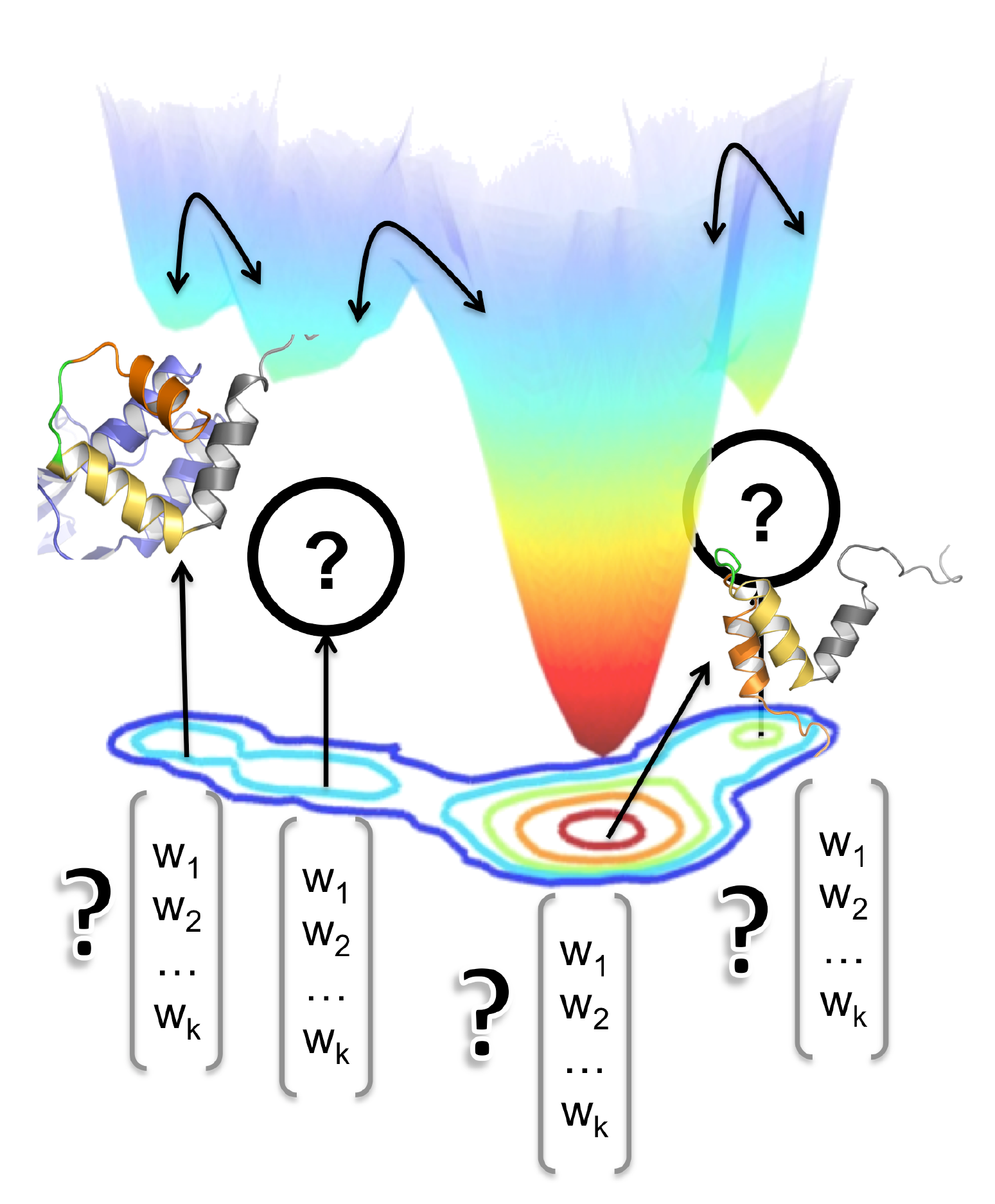}
\end{graphicalabstract}

\begin{highlights}
\item Recent successes of artificial intelligence (AI) and machine learning (ML) techniques can be leveraged to obtain quantitative insights into how intrinsically disordered proteins function. 
\item Review highlights the use of AI/ML techniques to characterize the intrinsic statistical coupling in IDP atomistic fluctuations involved in coupled folding and binding processes using linear, non-linear, and hybrid approaches. 
\item AI/ML methods can also be used to learn force-field parameters from long time-scale simulations as well as used to automatically coarse-grain IDP simulations. 
\item Bayesian inference methods in conjunction with AI/ML methods can be used to integrate sparse experimental observables to obtain a comprehensive picture of how IDPs function. 
\end{highlights}


\maketitle

\section{Introduction}\label{sec:Intro}
Our current understanding of protein structure-function relationships have been largely driven by the ability to visualize high-resolution three-dimensional (3D) structures of proteins with the aid of structure determination techniques including X-ray crystallography, nuclear magnetic resonance (NMR), and cryo-electron microscopy (cryo-EM)~\cite{Shin_2017}. These traditional structure determination techniques have often been supported with evidence from biochemical/biophysical methods to map out the functional consequences of perturbing protein structures through mutations and/or other modifications and for drug-discovery, protein design and other applications. However, the discovery of intrinsically disordered proteins (IDPs), and proteins with intrinsically disordered regions (IDRs) have challenged this traditional structure-function relationship paradigm~\cite{Uversky_2019}. In particular, IDPs/IDRs adapt their 3D structures exquisitely in response to their substrates as well as post-translational modifications (such as phosphorylation) and/or based on other physiological conditions (such as pH, crowding, etc.) and can mediate context-specific functions within cells~\cite{Philipps_2020}. Indeed, IDPs/IDRs are known to be equally sensitive to perturbations to their primary sequence, where mutations can have devastating effects including misfolding, protein aggregation (e.g., Parkinsons, Alzheimers and other ``conformational diseases'') and dysregulation of signaling pathways (e.g., cancer, diabetes, cardiovascular diseases)~\cite{Ruan_2019}. Given their central role in mediating complex biological functions within cells, understanding the structure-function paradigm of IDPs/IDRs remains an important challenge for modern biophysics. 

\begin{figure*}
    \centering
    \includegraphics[scale=0.43]{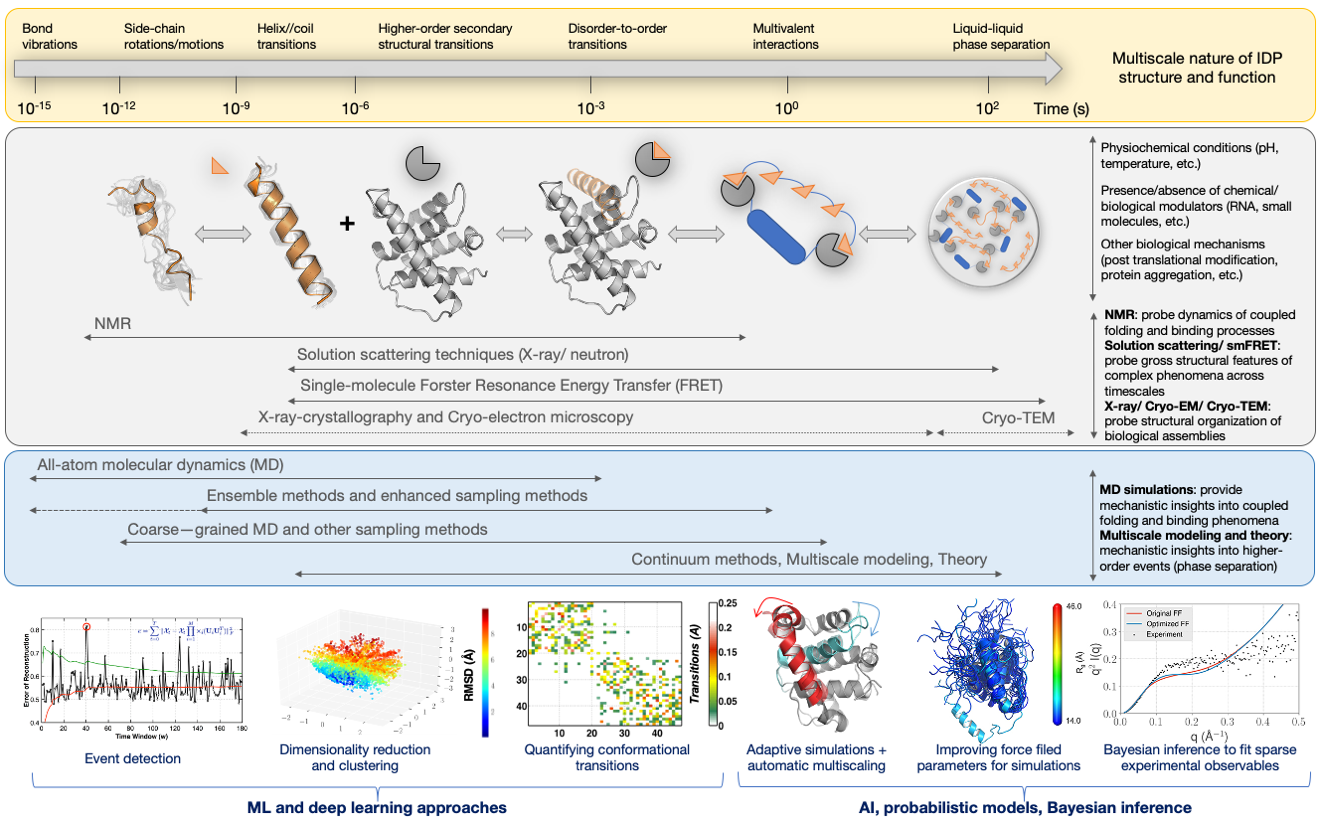}
    \caption{{\bf Role of AI/ML techniques in IDP/IDR biology.} Conformational fluctuations within IDPs occur at a wide range of time- (top panel) and length- (middle panel) scales. Further IDP systems are sensitive to physiological conditions, presence of biological modulators, and other mechanisms such as post-translational modifications. Solution scattering (X-ray/ neutron), smFRET and NMR techniques provide access to probe IDP fluctutations over a wide range of length- and time-scales; while X-ray and cryo-electron microscopy/ tomography provide access to static snapshots across longer length scales. It is notable that even within cryo-EM and TEM datasets, inherent limitations in resolution can result in a lot of the flexible regions missing, leading to the use of multiscale molecular simulations to fill in the gaps. However, even with improvements in enhanced/ adaptive sampling techniques, computational methods and computer hardware, it has been difficult to access details beyond $O(\mu$m$)$ length-scales and $O($ms$)$. We posit that AI/ML approaches will act as a `glue' that can enable integrating insights from simulations with experiments while providing a platform to interpret mechanisms of IDP/IDR function.}
    \label{fig:overview}
\end{figure*}

The remarkable plasticity of IDPs/IDRs is enabled by their ability to undergo folding upon binding -- one of the key mechanistic processes whereby an IDP/IDR adopts distinct secondary or even tertiary structure upon binding to a specific substrate (Fig. \ref{fig:overview}). This coupled folding and binding processes occur at diverse length- and time-scales -- beginning with finer conformational changes involving partial folding within IDR segments (e.g., helix-coil transitions) to disorder-to-order conformational transitions (e.g., formation of $\alpha-$helix) upon binding to a particular substrate~\cite{Majumdar_2019}. These local interactions can then drive the formation of higher-order interactions, whereby repeated ``segments'' of hydrophoic/polar amino-acid residues can transiently interact (albeit specifically) to their target substrates. These multivalent interactions in turn lead to coacervation or liquid-liquid phase separation (LLPS), which has important biological implications, including compartmentalization (e.g., membraneless organelles)~\cite{SCHULER202066}. One of the key challenges then is to elucidate the mechanisms by which IDPs undergo coupled folding and binding processes leading to such diverse functions. 

Although there has been tremendous progress in using traditional structure determination techniques in extending the length- and time-scales for studying IDPs~\cite{Xieeaax5560}, these techniques alone cannot fully describe the range of conformational flexibility of IDPs/IDRs. Further, given the intrinsic limitations in the length- and time-scales that these techniques can access, often multiple experiments are needed to probe the mechanisms by which IDPs/IDRs function, leading to a piecemeal approach in interpreting IDPs/IDRs ensembles~\cite{Rout_2019}. Molecular dynamics (MD) simulations, either via all-atom simulations or enhanced sampling techniques or multiscale coarse-grained methods provide a much needed `boost' in terms of sampling IDP conformational landscapes, allowing one to obtain insights into complex phenomena such as LLPS~\cite{DIGNON201992}. Synergy between experiments and simulations have been quite successful in quantitatively probing how  IDPs/IDRs function; however, such studies find it challenging when different experiments provide seemingly conflicting evidence that are not necessarily explained by simulations~\cite{ORIOLI2020123}.  

\paragraph{Motivating the need for AI/ML approaches in integrative IDP structural biology.} Advances in machine learning (ML) and artificial intelligence (AI) techniques have recently made strides in a number of scientific disciplines including molecular biophysics~\cite{NOE202077}. We posit that AI/ML techniques  can effectively act as a `glue' to integrate disparate sources of experimental and simulation data and to infer functional mechanisms of IDP/IDRs. In this review, we include a broad definition of how AI/ML methods are applied, where traditional statistical inference methods can be combined with methods that include neural networks. We examine how AI/ML techniques are being utilized in addressing the aforementioned challenges in IDP integrative structural biology, namely: (1) characterizing the conformational heterogeneity of IDP ensembles (Sec. \ref{sec:ML4ED}),  (2) multiscaling (length- and time-scales) IDP ensembles to model emergent phenomena such as LLPS (Sec. \ref{sec:ML4CG}), and (3) integration of sparse experimental observations with simulations to infer mechanisms of IDP function (Sec. \ref{sec:SI4ED}). Our review seeks to complement recent developments in AI/ML applications geared towards protein folding/ dynamics~\cite{NOE202077}. Further, we seek to bridge these advances in the context of simulation techniques for studying emergent behavior~\cite{DIGNON201992}. We finally conclude with a perspective on how AI/ML techniques can be integral in elucidating structure-function relationships of IDP/IDRs (Sec. \ref{sec:Conclusions}).


\section{AI/ML for characterizing IDP ensembles}\label{sec:ML4ED}
The range of conformations that IDPs can adapt is primarily attributed to the distribution of amino-acid residues along their primary sequences, where the ratio of charged residues to hydrophobic residues gives rise to specific patterning enabling them to vary their secondary (tertiary, and supra-molecular) structures in solution~\cite{Miskei_2019,Horvath_2020}. Since sequence based approaches by themselves are not sufficient to fully characterize IDP conformational landscapes, MD (and/or Monte Carlo) simulations are widely used to probe mechanisms of their functions, typically accessing timescales ranging $O$(10-100 $\mu$s)~\cite{Robustelli_2020}.  


\paragraph{Dimensionality reduction methods to organize IDP conformational landscapes.} ML/AI methods are necessary to quantify the statistical dependencies in atomistic fluctuations to obtain biophysically-relevant low-dimensional representations spanned by IDP landscapes. Dimensionality reduction methods summarize IDP ensembles in terms of a small number of collective variables or latent dimensions, where projections of the conformations from the simulations capture significant  events along these dimensions~\cite{Tribello_2019}. These projections are  referred to as \emph{embeddings}, where each conformation is represented by the latent dimensions. An implicit requirement of these embedding techniques is that they group conformations in terms of biophysically-relevant observables (e.g., root-mean squared deviations/RMSD, radius of gyration/R$_g$). Most dimensionality reduction techniques are unsupervised -- they exploit the intrinsic statistical structure within the data to discover dependencies without the need for explicit labels (for e.g., within an IDP ensemble, there is no explicit notion of what constitutes a folded/ partially folded/ unfolded state made available to the ML algorithm). Dimensionality reduction techniques can leverage linear, non-linear, or hybrid methods to learn low-dimensional embeddings and here we provide a succinct summary of how they have been used to characterize IDP ensembles~\cite{Ceriotti_2019}. 

Principal component analysis (PCA) is one such linear embedding method widely popular in analyzing simulation trajectory datasets~\cite{Tribello_2019}. However, PCA and its derivative methods lack the ability to characterize conformational diversity purely based on covariance in positional fluctuations alone. One key observation from several MD simulations as well as experimentally determined IDP ensembles is that their positional fluctuations exhibit long-tail distributions --  a natural consequence of their ability to undergo large conformational fluctuations and access \emph{rare} states away from their mean positions. These anharmonic fluctuations within IDPs are posited to be functionally relevant, since such fluctuations enable them to access conformational states relevant for binding to their specific substrate. The anharmonicity also gives rise to non-orthogonal correlations between individual atoms/amino-acid residues (depending on the resolution at which the data is being analyzed)~\cite{Ramanathan_2012}.



ML techniques such as anharmonic conformational analysis (ANCA) provide a convenient framework to analyze IDP ensembles especially in the context of disorder-to-order transitions~\cite{Parvatikar_2018}. ANCA uses fourth-order statistics to describe the atomic fluctuations and summarizes the internal motions using a small number of dominant anharmonic modes. In a recent study, time-resolved ANCA was used to characterize  disorder-to-order transitions in the BCL2 homology 3 domain, BECN1 (BCL2-interacting coiled-coiled protein) as it binds to the murine $\gamma$-herpesvirus 68 (M11) B-cell lymphoma 2 (BCL2) protein~\cite{Ramanathan_2020}. This approach identified a small number of conformational states that acted as intermediates in enabling M11-BCL2 to undergo partial unfolding in response to BECN1 binding. It identified a network of hydrophobic interactions, some farther than 10~\AA~ from the BH3D binding cleft that underwent specific conformational changes upon binding. These interactions were validated using mutagenesis and isothermal calorimetry demonstrating that perturning the intrinsic anharmonicity within M11 can adversely affect both protein stability and BECN1 binding. 

\paragraph{Deep-learning methods in analyzing IDP ensembles.} Long-tailed fluctuations in IDP ensembles is a characteristic indicator of multiscale behavior (Fig. \ref{fig:overview}). Further, the linearity assumptions in PCA and ANCA can be limiting in extracting multiscale features from the conformational landscape, especially when such embeddings are non-trivial. Deep learning methods that leverage neural networks have proven to be successful in progressively extracting multiscale features from raw inputs~\cite{Bengio_2015}. 

Deep neural networks such as autoencoders employ an hourglass shaped architecture where data is compressed into a low-dimensional latent space in the early layers and then reconstructed back~\cite{Doersch_2016}. The latent space learns to capture most essential information required for accurate reconstruction in the original dimensional space. Variational autoencoders (VAE) is one such instantiation of autoencoders that enforce the latent space to be normally distributed. Several variations of the VAE neural network architecture have been used to characterize  latent representations from protein folding trajectories, such as variational dynamics encoder (VDE~\cite{Hernandez_2018}), variational approaches for Markov processes (VAMP)~\cite{Mardt_2018}, reweighted autoencoded variational Bayes for enhanced sampling (RAVE)~\cite{Ribeiro_2018}, and the convolutional variational autoencoder (CVAE)~\cite{Bhowmik_2018,Bozkurt_Varolg_ne__2020}. Although the conceptual use of the VAE is similar, their  implementations can vary based on the essential  features that they are used to learn. For example, within VDE, the loss function includes a term capturing the slowest processes in the simulation datasets, whereas the EncoderMap~\cite{Lemke_2020A} utilizes a loss term that captures the proximity of conformations in the free-energy landscape. Complementary to these approaches, recurrent neural networks (RNNs) can serve as effective methods to learn time-dependent embeddings from MD simulations. RNNs, which are used extensively in natural language processing and image processing applications can be used to embed MD simulations to capture Boltzmann statistics from the system but also accurately reproduce the kinetics across multiple timescales~\cite{Tsai_2020}. Another approach by Noe and colleagues used a deep learning approach that is trained on a potential energy function and builds a generative model for conformational ensembles that respects Boltzmann statistics~\cite{Noe_2019Boltzmann}. The uniqueness of this approach is that it is `one-shot', meaning that it does not need any reaction coordinates and can produce unbiased samples, circumventing the expensive aspects of MD/Monte-Carlo simulations. 

IDRs often function as linkers between several folded domains (in multidomain proteins). This gives rise to an exponential number of states that they can sample, making it further challenging to characterize such complex landscapes. Dynamic graphical models (DGM) propose to address this problem by considering multidomain proteins as assemblies of coupled subsystems where each system is governed by the states it can access as well as the states its neighbors can access~\cite{Olsson15001}. Although DGMs use fewer parameters than their deep learning counterparts, it is difficult to incorporate prior experimental knowledge and recover atomistic configurations from its encoded representations. 
 

\section{AI/ML for multiscale simulations of IDP ensembles}\label{sec:ML4CG}
In the previous section, we described some of the recent developments applying ML approaches to characterize folding conformational landscapes. In this section, we examine how AI/ML methods can (1) inform efficient sampling of their conformational landscapes and (2) enable multiscale simulations of emergent phenomena such as LLPS.  

\paragraph{Determining reaction coordinates and enabling efficient sampling of IDP conformational landscapes. } The latent representations learned from MD simulations provide information relevant to reaction coordinates (RCs; also referred to as collective variables, or order parameters) that correspond to conformational changes along biophysically relevant observables (for e.g., $R_g$ values, or helicity, etc.). In a recent paper, Romero and colleagues demonstrated that the CVAE-learned embeddings can be used to cluster conformations from long time-scale simulations of the lysosomal enzyzme glucocerebrosidase-1 (GCase) and its facilitator protein saposin C (SAPC) along several reaction coordinates~\cite{Romero_2019}. The proposed conformational changes along the CVAE-determined RCs provided insights into key loop movements at the entrance of the substrate-binding site within GCase that are stabilized by direct interactions with SAPC. Note that this approach only used the raw simulation trajectories of GCase to infer the RCs and did not use any prior information (such as distance between residues or other features within GCase or SAPC).  Similar insights can be drawn from other approaches as well~\cite{rydzewski2020multiscale,Smith_2020,Fakharzadeh_2020}; however, the consequences of selecting a particular method versus what RCs they extract, and how they represent interpretable (biophysically meaningful) RCs remains an open question. 

RCs extracted from the analyses of MD simulations can be used to drive additional sampling of the conformational landscape. This is the basis for many adaptive and enhanced sampling approaches~\cite{Kasson_2018}. 
Techniques such as variational enhanced sampling (VES)~\cite{Bonati17641}, VAMPnets~\cite{Mardt_2018}, and RAVE~\cite{Ribeiro_2018} already include approaches for enhanced sampling. Both VES and VAMPnets utilize the variational approximation to enhance the sampling based on some set of reaction coordinates that can be determined by analyzing the MD simulations (see Sec. \ref{sec:ML4ED}). However, RAVE utilizes the predictive information bottleneck principle as an RC, where it can predict the most likely future trajectory  given a molecule's past trajectory. This principle, combined with the estimates for the most informative RCs (automatically determined from the information gain associated with sampling along subsets of RCs), the associated metastable states and equilibrium properties provides simultaneous access to uncover the unbiased kinetics for moving between different metastable states~\cite{Ribeiro_2019}. 

Generative adversarial networks (GAN)~\cite{goodfellow2016nips} have also been used for enhanced sampling, where on-the-fly training is used to modify the potential energy surface in order to drive the system to a user-defined target distribution where the free-energy barrier is lowered. This approach, called targeted adversarial learning optimized sampling (TALOS) uses MD simulations (for `generating' protein conformations) and a discriminator (differentiate samples generated by the biased sampler from those drawn from the desired target distribution) to automatically guide the sampling process~\cite{Zhang_2019}. This approach is inspired from actor-critic reinforcement learning ideas and is complementary to approaches such as reinforcement based adaptive sampling (REAP)~\cite{Shamsi_2018}. 


While AI/ML-driven MD simulations have been demonstrated for smaller peptide/protein systems, there is a need for effective middleware that can orchestrate complex workflows and manage resources efficiently~\cite{Kasson_2018}. Conventional (non deep learning) ML approaches take perhaps between a few seconds to may be a couple of hours to run and can easily be run concurrently with MD simulation jobs as long as the data is made available for analysis.  Training deep learning models on the other hand, can potentially take several hours (and even days) similar to the same timeline as MD simulations, which means resource management and scheduling has to be managed to make use of available compute time effectively. To address these issues, DeepDriveMD~\cite{Lee_2019} couples the CVAE~\cite{Bhowmik_2018} with adaptive MD simulations to accelerate folding of small proteins (up to 45 residues) on emerging supercomputers. DeepDriveMD's adaptive protocol could accelerate the sampling by at least 2.3x compared to traditional approaches. The adaptive sampling protocols used within simulation frameworks can be cast more generally as an optimization problem for balancing the cost of exploration (i.e., searching the IDP landscape) versus exploitation (i.e., utilizing existing knowledge to accelerate the search). The  AdaptiveBandit~\cite{Perez_2020} technique uses a reinforcement learning based approach where an action-value function and an upper confidence bound selection algorithm allows for substantial improvement of the sampling strategy.  

\paragraph{AI/ML approaches for learning force-field parameters and multiscale approaches.} Sampling IDP landscapes implies the need to access a wide range of conformations, even those with relatively low probabilities. While enhanced and adaptive sampling techniques provide an opportunity to access such low-probability conformational states, the timescales that simulations can access is still limited~\cite{Bhattacharya_2019}. Another potential challenge that limits the scale of sampling IDP/IDR landscape arises from the force field parameters used for these simulations. Several recent advances in force field parameter development do address these limitations specifically for IDP/IDR systems (see ~\cite{Zerze_2019,Yang_2019,Choi_2019}); however, artifacts related to how they are parameterized and how they end up capturing interfacial dynamics between IDPs and water (or other solvent conditions) still affect the overall quality of sampling~\cite{Zapletal_2020,Best_2020}. 

A complementary approach to this strategy is to use ML/AI to iteratively fit and refine force-field parameters in a data-driven fashion. One such approach, called ForceBalance-SAS~\cite{Demerdash_2019} (1) uses an initial `best' set of parameters, (2) computes ensemble averaged small-angle scattering intensities from MD simulations, (3) measures the residual with respect to experimental data, along with the gradient and Hessian of the residual, and (4) optimizes this fitting process from (1-3) iteratively until convergence criteria are achieved. This process continues with the newly updated set of parameters and simulations, completing the cycle. ForceBalance-SAS can optimize parameters for IDPs with varying molecular weight and different charge-hydrophobicity characteristics, albeit in a system-specific manner. While ForceBalance-SAS fits to the global small-angle scattering profiles, the force field parameters also resulted in better agreements with NMR chemical shifts (local observables). Further, the learned parameters could be transferred and applied to other systems partially (for shorter time-scale simulations). 

For simulating emergent phenomena such as LLPS,  coarse-graining is an essential step for making simulations tractable. While there are many approaches to coarse-grain simulations for LLPS (see review by Mittal and colleagues~\cite{DIGNON201992}), ML/AI approaches can aid in the development of data-driven representations from all-atom simulations for parameters needed at the coarse-grained resolution. One recent approach called lattice simulation engine for sticker and spacer interactions (LASSI) utilizes Boltzmann inversion, non-linear regression and a Gaussian process Bayesian optimization approach to parameterize the coarse-grained method for modeling sequence-specific phase-behaviors~\cite{Choi_2019_LASSI}. Additionally, force-field parameters simulating sequence-specific phase behavior could be enabled by an approach such as CAMELOT~\cite{Ruff_2015}.

Deep learning approaches can also be used to automatically infer coarse-grained representations from all-atom simulations~\cite{Zhang_2018A,Wang_2019}. Advances in graph neural networks are aiding the development of  accurate coarse-grained force field parameters~\cite{husic2020coarse}. It however remains to be seen how these approaches can be in turn generalized for IDP systems\cite{Noe_2020}. Similarly, the Multiscale Machine-learned Modeling Infrastructure (MuMMI)~\cite{Natale_2019} was developed to couple a continuum model with coarse-grained MD simulations using ML approaches to characterize how the oncogene RAS interacts with complex biological membranes. Complementary to this approach, adversarial autoencoders were coupled to multi-scale simulations of the severe acute respiratory coronavirus 2 (SARS-CoV-2) Spike protein in complex with the angiotensin-converting enzyme 2 (ACE2) receptor protein to probe the mechanisms of its infectivity~\cite{Casalino_2020}.   Automatic coarse-graining approaches using AI approaches can be really attractive for tuning the scale of coarse-graining that needs to be performed such that IDP/IDR landscapes can be adaptively sampled to obtain precise atomistic scale information about LLPS. Further, the ability to simulate self-consistent ensembles at multiple resolutions (continuum $\rightarrow$ coarse-grained $\rightarrow$ all-atom) will be critical for integrative structural biology applications in the context of combining information from diverse experimental techniques (see Sec. \ref{sec:SI4ED}.)   


\section{Statistical inference for integrating experimental data with simulations}\label{sec:SI4ED}

The previous sections outlined the use of AI/ML for characterizing IDP/IDR ensembles. But the true power of obtaining insights into the mechanisms of how IDPs function and how their functions can be exploited for therapeutic design~\cite{Ruan_2019}, novel material discovery~\cite{Dzuricky_2018}, and synthetic biology applications (e.g., membraneless organelles for transport)~\cite{Shin_2017} comes from the integration of theory and simulations with experimental data. The challenge with experimental data, however is that it can be noisy, sparse, and often provide only partial information when investigating a particular phenomenon~\cite{ORIOLI2020123}. For example, solution scattering data for IDPs are usually summarized using the scattering intensities against a coarse structural measure such as $R_g$,~\cite{Lipfert_2007} and in the case of single molecular Forster resonance energy transfer (sm-FRET) experiments, a set of distances is measured across the IDP structure~\cite{Metskas_2020}.  Simulations on the other hand, represent a full-scale system with all degrees of freedom (e.g., $3 \times N$, where N represents the individual atoms) implying a mismatch with the intrinsic dimensionality of experimental data. In such cases, how can one fit sparse experimental observables with simulation datasets? A second challenge arises when experimental datasets are unable to resolve flexible regions in a protein (e.g., cryo-electron microscopy)\cite{Lyumkis29032019}. Given that often such flexible regions hold key insights in terms of understanding ensembles of multi-domain proteins, simulations can fill in the gaps by providing probable states that these regions occupy. But the intrinsic gap in terms of timescales that can be accessed by simulations often ends up making it difficult to extract such information. Thus, AI approaches, augmented with Bayesian approaches can be quite helpful in bridging the gaps between experiments and simulations~\cite{Bottaro355,ORIOLI2020123}. 

There are two broad strategies for fitting simulation datasets with experiments. One strategy involves the use of unbiased simulations and then reweighting the generated ensembles using either maximum parsimony/ entropy approaches or with Bayesian strategies that uses information known from simulations as a \emph{prior} before the introduction of experimental observables. The complementary strategy involves the use of a biased simulations that are parameterized from experiments or using iterative approaches outlined in ~\cite{Demerdash_2019} to refine the force field parameters to sample the IDP landscape of interest. Similarly, integrated experimental and computational simulations are also being used to understand energetics of interactions between an IDP and its binding partner~\cite{Zou_2019}. Recent work by Lincoff and colleagues~\cite{Lincoff_2020} also extends the experimental inferential structure determination using a Bayesian formulation that calculates the maximum log-likelihodd of a conformational ensemble by accounting for the uncertainties across a variety of experimental data and back-calculation models. A similar integrated modeling approach by Gomes and colleagues~\cite{Gomes_2020} demonstrated how conformational restraints imposed using NMR, SAXS, and sm-FRET approaches could reach agreement in the ensembles of Sic1 and phosphorylated Sic1. 

\section{Challenges and outlook}\label{sec:Conclusions}
From quantitatively probing the complex conformational landscapes of IDPs to identifying disorder-to-order transitions or modeling emergent phenomena such as LLPS, ML/AI approaches are proving to be an indispensable tool for both experimental biophysicists as well as modelers. However, ML/AI approaches for MD simulations still face some challenges that need to be addressed. 

Current AI/ML applications (barring a few like~\cite{Parvatikar_2018,Noe_2019Boltzmann}) tend to use fitting procedures in a blind manner, without much physical bearing, or paying attention to the underlying statistical physics of the system of interest. The resulting fitting procedure can end up overfitting and may not generalize to fully leverage the power of ML/AI in other domains~\cite{Pant_2020,Goolsby_2020}. In particular, transferring a ML/AI model learned across simulations can be challenging. ML/AI methods may also get stuck in regimes that are not entirely physical -- leading to issues in how appropriate (weighted) sampling can be achieved. Although techniques such as cross-validation and regularization alleviate these problems, there is a need to develop rigorous statistical techniques as well as interactive  tools that can assess the performance of ML/AI models. A second challenge arises when force field parameters are designed using ML/AI. Here, the challenge is in maintaining control over the versions of the force-field designed by ML/AI approaches -- where initial conditions or datasets used for training, the inherent stochastic nature of how deep learning approaches work, and even program implementations (differences between how TensorFlow and PyTorch modules are implemented) -- can result in highly divergent results, even if the physically represented parameters may in fact lie reasonably within the same range.  While the ML/AI community already does similar activities through rigorous benchmarking applications~\cite{Mattson_2020}, a similar effort from the IDP/IDR community is needed to ensure robustness, reusability, and reproducibility of models across multiple studies. Efforts such as the IDP ensemble~\cite{Varadi2015} database provide for such an opportunity; however, there is a need for the community-wide engagement to assess these intrinsic issues.

Further, many of the ML/AI results are considered \emph{black box}, meaning that it is difficult to reason how the ML/AI model made its inference. Even though there have been some advances in enabling  interactivity with the outputs from the ML/AI models~\cite{Chae_2019}, there is still the challenge of making it interactive when large datasets are streamed. Developments in interactive data analysis and virtual reality can aid this, although significant developments are needed to make these approaches practical for emerging datasets.

Finally, computational infrastructure to support ML/AI workflows in concert with simulations has been a long-standing challenge~\cite{fox2019learning}. Traditional approaches run MD simulations continuously, store these large datasets and eventually analyze them with ML/AI methods. However, in the Exascale computing era, such approaches will become infeasible as the sheer volume of data generated by these machines can far exceed the capabilities of analyses that needs to be done (and occasionally, computing resources for AI/ML can exceed that of simulations).  Approaches such as DeepDriveMD~\cite{Lee_2019} are examining such emerging needs of complex workflows; however, we believe there is much research that needs to be done in order to understand how AI/ML workloads will interact with future simulation workloads. With newer developments in AI techniques, there is an opportunity to accelerate our understanding of how IDPs play a role in disease, developing novel means to design small-molecule inhibitors, designing new bio-materials, and engineering self-assembling systems for synthetic biology applications. We believe these represent exciting opportunities for the future.

\section*{Acknowledgements}
A.R. thanks Anda Trifan for assistance in editing and proof-reading the manuscript. This research was supported by Argonne Laboratory Directed Research and Development Computing Expedition project (A.R.) and NIH/NIGMS GM105978 (S.C.C.). 

\bibliographystyle{cas-model2-names}

\bibliography{main}


\end{document}